\documentstyle[aps,12pt,floats,psfig]{revtex}
\tighten
\begin{document}

\preprint{}

\title{Properties of Strange Hadronic Matter in Bulk and in Finite Systems}

\author{J\"urgen Schaffner-Bielich}

\address{RIKEN BNL Research Center, Brookhaven National Laboratory,\\
Upton, New York 11973}

\author{Avraham Gal}

\address{Racah Institute of Physics, The Hebrew University, Jerusalem 91904,
Israel}

\maketitle

\begin{abstract}

        The hyperon-hyperon potentials due to a recent SU(3) Nijmegen
        soft-core potential model are incorporated within
        a relativistic mean field calculation of strange hadronic matter.
        We find considerably higher binding energy in bulk matter compared to
        several recent calculations which constrain the composition of
        matter. For small strangeness fractions
        ($f_{S} \lesssim 1$), matter is dominated
        by $N \Lambda \Xi$ composition and the calculated binding energy
        closely follows that calculated by using the hyperon potentials of our
        previous calculations. For larger strangeness fractions ($f_{S}
        \gtrsim 1$), the calculated
        binding energy increases substantially beyond any previous calculation
        due to a phase transition into $N \Sigma \Xi$ dominated matter.
        We also compare bulk matter calculations with
        finite system calculations, again highlighting the consequences of
        reducing the Coulomb destabilizing effects in finite strange systems.

\end{abstract}

\section{Introduction}

Bodmer and Witten independently highlighted the idea that strange quark
matter, with roughly
equal composition of $u$, $d$ and $s$ quarks leading to a strangeness
fraction $f_{S} = -S/A \approx 1$ and a charge fraction $f_{Q} = Z/A
\approx 0$, might provide the absolutely stable form of matter
\cite{Bodmer71,Wit84}. Metastable strange quark matter has been studied by
Chin and Kerman \cite{Chin79}. Jaffe and collaborators \cite{Jaf84,Jaf87}
subsequently charted the various scenarios possible for the stability
of strange quark matter, from absolute stability down to metastability
due to weak decays. Finite strange quark systems, so called strangelets,
have also been considered \cite{Jaf84,Jaf93}. For a recent review of
theoretical studies and experimental searches for strangelets,
see Refs.\ \cite{Scha98,Nagle99}.

Less advertised, perhaps, is the observation made  in our previous work
\cite{Scha93,Scha94} that metastable strange
systems with similar properties, i.e. $f_{S} \sim 1$ and $f_{Q} \sim 0$, 
might also exist in the hadronic basis at moderate values of
density, between twice and three times nuclear matter density. These
strange systems are made out of nucleons ($N$), lambda ($\Lambda$) and
cascade ($\Xi$) hyperons. The metastability of these strange
hadronic systems was established by extending relativistic mean field
(RMF) calculations from ordinary nuclei ($f_{S} = 0$) to
multi-strange nuclei with $f_{S}\not= 0$.
Although the detailed pattern of metastability, as well as
the actual values of the binding energy, depend specifically on
the partly unknown hyperon potentials assumed in dense matter, the
predicted phenomenon of metastability turned out to be robust in
these calculations \cite{Scha94,Bal94}.

Quite recently, Stoks and Lee \cite{Stoks99b} have challenged the
generality of the above results for strange hadronic systems. These
authors constructed $G$ matrices for coupled baryon-baryon channels,
using an SU(3) extension \cite{Stoks99a} of the Nijmegen
soft-core NSC97 potentials \cite{RSY99} from the $S = 0, -1$ sector
(to which data these potentials have been fitted) into the unexplored
$S = -2, -3, -4$ sector. These $G$ matrices were then employed within a
Brueckner-Hartree-Fock (BHF) calculation of strange hadronic matter (SHM)
in bulk. The results showed that $N \Lambda \Xi$ systems are only
loosely bound, and that charge-neutral strangeness-rich hadronic
systems are unlikely to exist in nature in metastable form, in stark
contrast to our earlier findings \cite{Scha93,Scha94}.

This vast difference in the predictions for the metastability and
binding of SHM between following a BHF methodology, which uses an SU(3)
extrapolated form of the NSC97 baryon-baryon potentials, and following
a RMF methodology, which is based on mean fields designed to mimic the
consequences of the Nijmegen hard-core potential model D \cite{Nag77}, has
prompted us to investigate possible origins of it. In this work we
present calculational evidence for the incompleteness of the procedure
applied by Stoks and Lee \cite{Stoks99b}. We do so by reproducing
qualitatively their results for the instability and weak binding of $N
\Lambda \Xi$ matter in bulk, within a {\it constrained} RMF
calculation in which the mean fields are now designed to mimic the
consequences of the NSC97 model used by Stoks and Lee. The
constraints imposed by us, as a check, are identical with those
imposed by these authors for the composition of SHM (see Fig. 4 of
Ref. \cite{Stoks99b}). We argue that this is not the right way to
identify minimum-energy equilibrium configurations for SHM. Indeed,
doing the {\it unconstrained} RMF calculation with the same
NSC97-inspired mean fields, we find qualitatively good agreement, for
$f_{S} \lesssim 1$ where the bulk matter is $N \Lambda
\Xi$ dominated, between these new results and our old results in
model 2 \cite{Scha94}. For $f_{S} \gtrsim 1$, the
new unconstrained calculation results in considerably higher binding
energies than ever calculated for SHM, due to a phase transition into $N
\Sigma \Xi$ dominated matter.

The paper is organized as follows. In section II we describe
the methodology of finding equilibrium configurations within the RMF
formalism, and the input mean fields entering the new RMF
calculations. Section III includes the results of these new
calculations for {\it bulk} SHM, as well as for {\it finite} multi-strange
systems for which BHF calculations have not been done to date. The role of
the Coulomb interaction in stabilizing charge-neutral strange systems
is highlighted. Our results are summarized and discussed in
section IV, where we also comment on the applicability of the SU(3)-extended
NSC97 potential.

\section{Methodology and input}

We adopt the Relativistic Mean Field (RMF) Model to describe strange hadronic
matter (SHM) in bulk and for finite systems of nucleons and hyperons.
The model is an effective model where the parameters are adjusted to the known
properties of nuclei and hypernuclei. We include in our extended RMF model all
the $1/2^+$ baryons of the lowest SU(3) flavor octet, as well as
hidden-strangeness meson exchange to allow for possibly strong hyperon-hyperon
($YY$) interactions.
Here we use model 1 and model 2 of Ref.\ \cite{Scha93}. The basic ingredients
of these models are the octet baryons matrix $B$, the matrices $V^{(8)}_\mu$
and $V^{(1)}_\mu$ of the vector meson octet and singlet, respectively, and
the two scalar mesons $\sigma$ and $\sigma^*$.
In addition, a Coulomb term is included in finite system calculations.
The Lagrangian is given as
\begin{eqnarray}
{\cal L} &=&
\mbox{Tr} \bar B (i \gamma^\mu\partial_\mu
- g_{\sigma B} \sigma - g_{\sigma^* B} \sigma^*
- m_B) B
\cr &&
- \frac{1}{2}\left(\partial^\mu \sigma \partial_\mu \sigma
- m_\sigma^2 \sigma^2\right)
- \frac{b}{3}\sigma^3 - \frac{c}{4}\sigma^4
- \frac{1}{2}\left(\partial_\nu\sigma^*\partial^\nu\sigma^*
- m_{\sigma^*}^2{\sigma^*}^2\right)
\cr &&
- g^{(8)}_{v} \left( \alpha \mbox{Tr} \bar B \gamma^\mu \left[V^{(8)}_\mu,
    B\right] + \left(1-\alpha\right) \mbox{Tr} \bar B \gamma^\mu
  \left\{V^{(8)}_\mu,B\right\} \right)
- g^{(1)}_{v} \mbox{Tr} \bar B \gamma^\mu B \cdot \mbox{Tr} V^{(1)}_\mu
\cr &&
- \frac{1}{4} \mbox{Tr} V^{\dagger}_{\mu\nu}V^{\mu\nu}
+ \frac{1}{2}  \mbox{Tr} m_v^2 V^{\dagger}_{\mu} V^\mu
+ \frac{1}{4} d (\omega_\mu \omega^\mu)^2
\quad .
\end{eqnarray}
Here, both scalar fields are treated as singlets. The octet vector fields can
be coupled in two ways, either antisymmetric (F-type with $\alpha=1$) or
symmetric (D-type with $\alpha=0$).
In the mean-field approximation, only the $\omega$, $\rho$, and $\phi$ vector
mesons remain operative. Their coupling constants to the baryon fields can be
related by SU(3) symmetry \cite{Dover84}. By assuming ideal mixing of the
vector mesons (i.e.\ the $\phi$ is a purely $s\bar s$ state), pure F-type
coupling ($\alpha=1$), and that the nucleon does
not couple to the $\phi$, one recovers the SU(6) relations of the simple quark
model
\begin{eqnarray}
&& \frac{1}{3}g_{\omega N} = \frac{1}{2} g_{\omega\Lambda}
= \frac{1}{2} g_{\omega\Sigma} =  g_{\omega\Xi}
\cr &&
g_{\rho N} = \frac{1}{2} g_{\rho\Sigma} =  g_{\rho\Xi} \; ,
\quad g_{\rho\Lambda} = 0
\cr &&
2g_{\phi\Lambda} = 2g_{\phi\Sigma} = g_{\phi\Xi} = - \frac{2\sqrt{2}}{3}
g_{\omega N} \; , \quad g_{\phi N} = 0 \quad .
\label{eq:su6}
\end{eqnarray}
Here, the constraint $3g_{\rho N} = g_{\omega N}$ is relaxed to allow
the isovector coupling constant to be fixed by the isospin dependence of
nuclear binding energies.
The quark model is not used for the scalar coupling constants, rather they
are determined by adjusting to nuclear and hypernuclear properties.
Self-interaction terms for the scalar field $\sigma$ and the vector field
$\omega$ are also included in the model. In the following, we use the
parameter set TM1 of Ref.\ \cite{Suga94} where the parameters were taken
from a fit to properties of spherical nuclei.
The remaining scalar $\sigma$ coupling constants for
the hyperons are chosen to give reasonable hyperon potentials in saturated
nuclear matter:
\begin{equation}
U_\Lambda^{(N)}(\rho_0) = -30 \mbox{ MeV },\quad
U_\Sigma^{(N)}(\rho_0) = +30 \mbox{ MeV }, \mbox{ and} \quad
U_\Xi^{(N)}(\rho_0) = -18 \mbox{ MeV }.
\label{eq:hyppot}
\end{equation}
Note, that these relativistic potentials are $(10-20) \%$ stronger than
the corresponding nonrelativistic values. The values for $\Sigma$ and $\Xi$
hyperons differ from the previous choice of Refs.\ \cite{Scha93,Scha94},
reflecting recent developments in hypernuclear physics which are briefly
recorded below.

For the $\Sigma$ nuclear interaction, the most updated analysis of $\Sigma^-$
atomic data indicates a repulsive isoscalar potential in the interior of nuclei
\cite{Mares95} which is compatible with the absence of bound-state or continuum
peaks in a recent search for $\Sigma$ hypernuclei \cite{Bart99}. In fact, the
only $\Sigma$ hypernuclear bound state found so far is $^4_\Sigma$He
\cite{Hayano89,Nagae98}, where the binding results from the strong isovector
component of the $\Sigma$ nuclear interaction. These statements are supported
by several recent calculations \cite{Dab00,Harada00}. The precise magnitude of
the depth of the $\Sigma$ nuclear potential is of little importance
to our investigations. It turns out that the $\Sigma$ hyperon will not
appear anyway in the bulk matter calculation, or in the finite system
calculations within models 1 and 2, unless its hadronic interactions are
exceptionally strong, so as to block the release of about 75 MeV in the
free-space $\Sigma B \to \Lambda B$ strong-interaction conversion.

For the $\Xi$ nuclear interaction, measurements of the final-state interaction
of $\Xi$ hyperons produced in the ($K^{-}, K^+$) reaction on $^{12}$C in
experiments E224 at KEK \cite{Fukuda98} and E885 at the AGS \cite{Khaustov99}
indicate a nonrelativistic potential $U_{\Xi, \rm nr}^{(N)}$ of about $-16$
and $-14$ MeV or less, respectively. Below we will actually vary the value for
$U_\Xi^{(N)}$ to check its effect on the binding energy of SHM.

The hyperon ($Y$) potentials $U^{(Y')}_Y$ in hyperon ($Y'$) matter,
in the absence of direct experimental data, depend to a large extent on
the assumptions made on the underlying $YY$ interactions.
In model 1, which does not use $\sigma^*$ and $\phi$ exchanges, the potentials
$U^{(Y')}_Y$ are rather weak, less than 10 MeV deep. The exchange of these
hidden-strangeness mesons is included in model 2, where the $\sigma^*$
coupling to hyperons is adjusted so that the potential of a single hyperon,
embedded in a bath of $\Xi$ matter at nuclear saturation density $\rho_0$,
becomes
\begin{equation}
U^{(\Xi)}_{\Xi}(\rho_0) = U^{(\Xi)}_{\Lambda}(\rho_0) = -40 \mbox{ MeV} \quad ,
\label{eq:hypbath}
\end{equation}
in accordance with the attractive $YY$ interactions of the Nijmegen
potential model D \cite{Scha94}. The resulting $U_\Lambda^{(\Lambda)}$
is about $-20$ MeV, considerably more attractive than in model 1.
Indeed the few double $\Lambda$
hypernuclear events observed so far in emulsion require a relatively
strong $\Lambda\Lambda$ attractive interaction \cite{Dover91}, which
lends support to model 2 over model 1, but the actual situation for the
other, unknown, $YY'$ channels could prove more complex than allowed for by
either model. All that may be said at present is that, as far as the
$\Lambda\Lambda$ interaction strength is concerned, model 2 is a more
realistic one than model 1.

Since there appears some confusion in the recent literature
\cite{Stoks99b,Vidana99} regarding how to calculate self consistently
the properties of SHM in bulk, we will ponder on the thermodynamically
consistent methodology in more detail. 
As we will demonstrate in the following, a major property of SHM within
the SU(3)-extended NSC97 model might have been overlooked in these works.
Here we focus on the
thermodynamically correct treatment in the RMF approximation.
The extension to BHF calculations is then straightforward. Very recently,
thermodynamically consistent BHF calculations of $\beta$-stable strange matter
in neutron stars 
have been performed by Baldo {\it et~al.} \cite{Baldo99}, using the
NSC89 model \cite{NSC89} for the $YN$ interactions, and by Vida\~{n}a
{\it et~al.} \cite{Vidana00}, using the SU(3)-extended NSC97 model
\cite{Stoks99a} for the $YN$ and $YY$ interactions.

In general, we can describe the system by the grand-canonical
thermodynamic potential $\Omega$, which depends on the temperature $T$,
the volume $V$, and the independently conserved chemical potentials
$\mu_{\alpha}$. At $T$ = 0, the pressure is given by:
\begin{equation}
P({\mu_{\alpha}}) = - \Omega({\mu_{\alpha}},T=0) / V \quad .
\end{equation}
For SHM in bulk,
since the isospin dependence is usually suppressed, there are just two conserved
charges in bulk which are the baryon number $B$ and the strangeness number $S$.
The chemical potentials of the individual baryons can be related to the
corresponding baryon chemical potential $\mu_B$ and strangeness chemical
potential $\mu_S$ by
\begin{equation}
\mu_i = B_i \cdot \mu_B + S_i \cdot \mu_S
\label{eq:chemeq}
\quad .
\end{equation}
This ensures that the system is in chemical equilibrium or, in other words,
that the strangeness and baryon numbers are conserved in all possible
strong-interaction reactions in the medium, such as
\begin{equation}
\Sigma + N \leftrightarrow \Lambda + N \quad
\Lambda + \Lambda \leftrightarrow \Xi + N \quad
\Lambda + \Xi \leftrightarrow \Sigma + \Xi \quad \dots
\end{equation}
The Hugenholtz -- van-Hove theorem relates the Fermi energy of each baryon to
its chemical potential in equilibrated matter
\begin{equation}
\mu_i = E_{F,i} = \sqrt{k_{F,i}^2 + {m_i^*}^2} + g_{\omega i} \omega_0
 + g_{\rho i} \rho_0 + g_{\phi i} \phi_0
\label{eq:HvH}
\quad .
\end{equation}
Here we used the energy-momentum relation of the mean-field approximation with
the effective mass $m_i^*=m_i+g_{\sigma i}\sigma + g_{\sigma^* i} \sigma^*$ for
the baryon species $i$. Note, that these potentials depend on the overall
composition of the matter, requiring thus a self-consistent calculation.
Equation (\ref{eq:HvH}) can be easily solved to
calculate the Fermi momenta $k_{F,i}$ and hence the number density of baryon $i$ for
given chemical potentials,
\begin{equation}
\rho_i = \gamma_i \frac{k_{F,i}^3}{6 \pi^2} \quad ,
\end{equation}
where $\gamma_i$ is the spin-isospin degeneracy factor. If the solution
results in an imaginary Fermi momentum, the particle is not present in the
system and the corresponding density is set to zero.
In Brueckner theory, one has to solve for an equation
of the form
\begin{equation}
\mu_i = E_i (k_{F,i}) = m_i + \frac{k_{F,i}^2}{2m_i} + \Re U_i (k_{F,i})
\end{equation}
since the potential is now momentum dependent.
It is apparent from the above procedure that no baryon species may be
ignored a priori, but one has to check for their appearance by calculating
the corresponding Fermi momentum. Therefore, any calculation of baryonic
matter with nucleons and $\Lambda$'s alone is bound to violate the condition
of chemical equilibrium Eq.\ (\ref{eq:chemeq}), since $\Sigma$ and
particularly $\Xi$ hyperons are likely to appear in strange hadronic systems
\cite{Bal94}. In that sense, multi-$\Lambda$ matter calculations as performed
in \cite{Ikeda85,MZ89,Rufa90,Barranco91,Lanskoy92,Zhang97,Schulze98}
are incomplete.
Furthermore, the Fermi momenta of different baryon species cannot be set
equal to each other, since this again violates the condition of chemical
equilibrium. If one arbitrarily sets certain baryon fractions to be equal to
each other, as e.g.\ done in \cite{Stoks99b}, the resulting system is not in
its energetically favored global minimum and the computed binding energies
will be underestimated. In addition, the overall pressure of the system will be
too low, resulting in a too soft equation of state; maximum masses of neutron stars
computed in this way will be underestimated.

The pressure and the energy density in the RMF model in bulk are given by
\begin{eqnarray}
P &=& - \frac{1}{2}m_\sigma^2 \sigma^2
- \frac{b}{3}\sigma^3 - \frac{c}{4}\sigma^4
- \frac{1}{2}m_{\sigma^*}^2 {\sigma^*}^2
+ \frac{1}{2}m_\omega^2 \omega_0^2 + \frac{1}{4} d \omega_0^4
+ \frac{1}{2}m_\phi^2 \phi_0^2
\cr && {}
+ \sum_{i=B,l} \frac{\gamma_i}{(2\pi)^3} \int_0^{k_{F,i}} d^3 k
\frac{k^2}{\sqrt{k^2 + {m^*_i}^2}}
\quad ,
\cr
\epsilon &=& \frac{1}{2}m_\sigma^2 \sigma^2
+ \frac{b}{3}\sigma^3 + \frac{c}{4}\sigma^4
+ \frac{1}{2}m_{\sigma^*}^2 {\sigma^*}^2
+ \frac{1}{2}m_\omega^2 \omega_0^2 + \frac{3}{4} d \omega_0^4
+ \frac{1}{2}m_\phi^2 \phi_0^2
\cr && {}
+ \sum_{i=B,l} \frac{\gamma_i}{(2\pi)^3} \int_0^{k_{F,i}} d^3 k
\sqrt{k^2 + {m^*_i}^2}
\quad ,
\end{eqnarray}
respectively.
The binding energy per baryon is then obtained by subtracting the
properly weighted combination of the rest masses from the energy
density of the system
\begin{equation}
E/A = \frac{1}{\rho_B} \left(\epsilon
   - \rho_N \cdot m_N - \rho_\Lambda \cdot m_\Lambda - \rho_\Sigma \cdot
   m_\Sigma - \rho_\Xi \cdot m_\Xi \right)
\quad .
\end{equation}
The meson fields are determined by their equations of motion (see e.g.\
\cite{SM96}). The particle densities are calculated using the
thermodynamically consistent formalism as outlined above.

For most purposes, one wishes to translate the dependence on the two chemical
potentials into their corresponding baryon and strangeness number density.
This is done using the expressions for overall baryon and strangeness number
conservation
\begin{mathletters}
\begin{eqnarray}
\rho_B &=& \sum_i B_i \cdot \rho_{V,i}
= \rho_N + \rho_\Lambda + \rho_\Sigma + \rho_\Xi \\
\rho_S &=& \sum_i (-S_i) \cdot \rho_{V,i}
= \rho_\Lambda + \rho_\Sigma + 2\rho_\Xi
\end{eqnarray}
\end{mathletters}
where $\rho_{V,i}$ is the vector density of baryon species $i$.

\begin{figure}[tbph]
\begin{center}
\leavevmode
\psfig{file=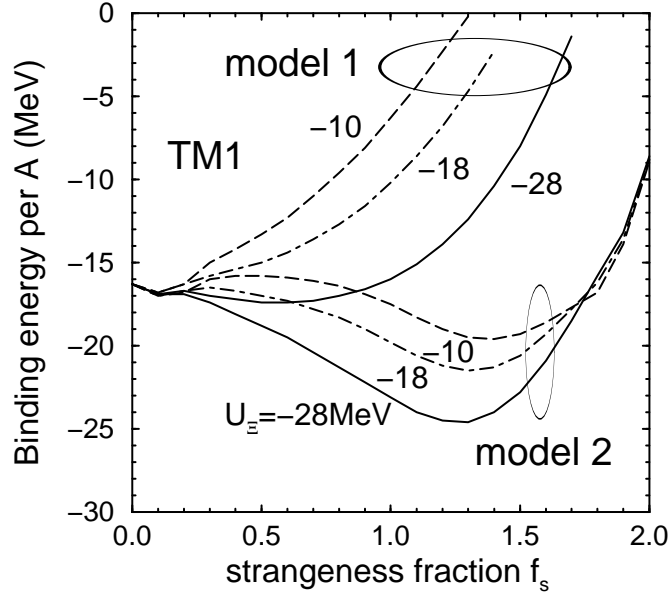,height=0.35\textheight}
\end{center}
\caption{Binding energy per baryon of SHM in models 1 and 2,
with different choices of the $\Xi$ potential in nuclear matter.}
\label{fig:eafsm3}
\end{figure}

A useful measure of the strangeness contents of the system is given
by the strangeness fraction:
\begin{equation}
  f_S = \frac{\rho_S}{\rho_B} =
\frac{\rho_\Lambda + \rho_\Sigma + 2\rho_\Xi}
{\rho_N + \rho_\Lambda + \rho_\Sigma + \rho_\Xi}
\quad.
\end{equation}

\section{Results}

Plots of calculated binding energy of SHM per baryon as function
of the strangeness fraction $f_S$ are shown in Fig.\ \ref{fig:eafsm3},
for different choices of the $\Xi$ nuclear potential denoted by $U_\Xi$,
in models 1 and 2 \cite{Scha93,Scha94}. For $f_S=0$, there are only nucleons in
the system and one gets the standard equation of state of nuclear matter as
function of baryon density. The equilibrium density of nuclear matter is
determined by minimizing the binding energy with respect to the baryon density.
The resulting minimum value of binding energy per nucleon is shown then at
$f_S=0$ in the plots of Fig.\ \ref{fig:eafsm3}. Next, we increase the
strangeness fraction from zero on, and the system of equations adjusts itself
at each fixed value of $f_S$ to find the
corresponding baryon densities ensuring chemical equilibrium (Eq.~(\ref{eq:chemeq})).
The minimum value of the binding energy per baryon as function of baryon density
at each fixed strangeness fraction is then plotted in Fig.\ \ref{fig:eafsm3}
for the corresponding value of $f_S$.
In this way, one gets the binding energy of SHM as function of the strangeness
fraction. It turns out that $\Sigma$ hyperons do not appear at any value of
$f_S$ in both models 1 and 2. To display the dependence on the $\Xi$ nuclear
potential we chose three different values, $U_{\Xi}=-10,-18,-28$ MeV. The
variation in the plots of model 1 is quite pronounced. For $U_\Xi=-28$ MeV,
the minimum is at a finite value $f_S=0.6$, with a binding energy per baryon
of $-17.4$ MeV. For shallower $\Xi$ potentials, this minimum disappears
and slightly strange matter with $f_S\approx 0.1$ is the most strongly bound configuration.
On the other hand, in model 2, varying $U_\Xi$ does not lead to drastic changes.
The minimum in the binding energy per baryon for $U_\Xi=-28$ MeV, at $f_S=1.3$
with $E/A=-24.6$ MeV, is shifted to $E/A=-21.5$ MeV for $U_\Xi=-18$ MeV and to
$E/A=-19.6$ MeV at a slightly higher value $f_S=1.4$ for $U_\Xi=-10$ MeV. The
reason is that in model 2 the minimum is generated by the $YY$ interactions
which have been adjusted according to Eq.\ (\ref{eq:hypbath}),
so that the binding energy curves in model 2 are not as much affected by
changing $U_\Xi$ as compared to the effect of this change in model 1. Note
that the constraint (\ref{eq:hypbath}) ensures that pure $\Xi$ matter
($f_S=2$) has the same binding energy, $E/A=-8.9$ MeV, in all three cases.
Pure $\Xi$ matter is always unbound in model 1 due to the missing
attraction in the $YY$ channels.

\begin{figure}[tbph]
\begin{center}
\leavevmode
\psfig{file=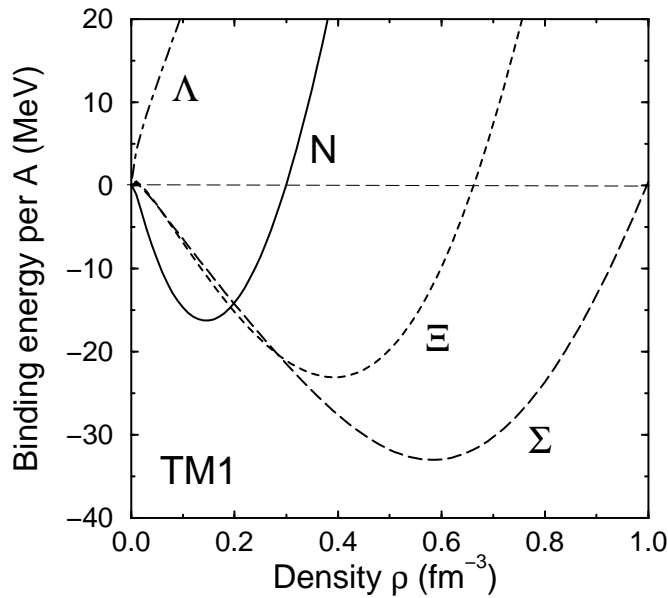,height=0.35\textheight}
\end{center}
\caption{Binding energy per nucleon ($N$) in nucleon matter, compared to the
  binding energy per hyperon ($\Lambda,\Sigma,\Xi$) in its own hyperonic matter.
  The hyperonic parameters were chosen to reproduce the binding energy minima
  of Fig. 2 in Ref.\ \protect\cite{Stoks99b}.}
\label{fig:earho_hyp}
\end{figure}

Substantial departures from the universality (Eq.\ (\ref{eq:hypbath})) assumed
in Refs.\ \cite{Scha93,Scha94} for the $YY$ interactions occur in the most
recent SU(3)-extension of the Nijmegen soft-core potential model NSC97
\cite{Stoks99a}. In particular, the $\Sigma\Sigma$ and $\Xi\Xi$ interactions
are predicted to be highly attractive in some channels, leading to bound states.
We wish to examine the consequences of this model in our RMF calculation of
SHM. The $YY$ interactions of Ref.\ \cite{Stoks99a} are implemented in our
calculation by adjusting the coupling constants of the $\sigma^*$ meson field
to reproduce qualitatively the hyperon binding energy curves shown in Fig.~2
of Ref.\ \cite{Stoks99b} for set NSC97f. All the other coupling constants
are held fixed, so that we still get the hyperon potentials of
Eq.\ (\ref{eq:hyppot}) in nuclear matter. The resulting binding energy curves,
of each baryon species $j$ in its own matter $B_j^{(j)}$,
are depicted in Fig.\ \ref{fig:earho_hyp} as function of density.
For nucleons, we again use the parameterization TM1 so as to get the correct
binding energy at the correct saturation density $\rho_0$.
Note that NSC97f does not reproduce the correct nuclear matter saturation point,
but gives a too shallow minimum at a too high density
(see Fig.~2 of Ref.\ \cite{Stoks99b}). No binding occurs for $\Lambda$
hyperons, and $B_\Lambda^{(\Lambda)}$ reaches $+20$ MeV already at rather
low density, $\rho=0.1$ fm$^{-3}$. This strong repulsive ``potential''
is due to the very weak underlying $\Lambda\Lambda$ interaction in the
extended NSC97f model which is incompatible with the fairly strong
$\Lambda\Lambda$ attraction necessary to explain the observed double
$\Lambda$ hypernuclear events (see \cite{Dover91,Aoki91} and references
therein). On the other hand, $\Sigma$ matter is deeply bound, by $-33$ MeV
per baryon at $\rho =0.58$ fm$^{-3}$ which is twice as deep as ordinary
nuclear matter, and $\Xi$ matter has a binding energy of $-23$ MeV per baryon
at $\rho=0.39$ fm$^{-3}$.
It is clear from Fig.\ \ref{fig:earho_hyp} that a mixture of $\Sigma$ and
$\Xi$ matter must be very deeply bound too, unless there is an overwhelmingly
repulsive interaction between $\Sigma$ and $\Xi$ hyperons. Actually, the
interaction between $\Sigma$ and $\Xi$ hyperons is the most attractive one in
the extended NSC97f model, giving rise to  the deepest bound $\Sigma\Xi$
dibaryon state \cite{Stoks99a}.
Independently, increasing the number of degrees of freedom will also result in
a more deeply bound state. This is the case, for example, when going from
unbound neutron matter ($\gamma=2$) to bound nuclear matter ($\gamma=4$).
Therefore, one expects that $\Sigma\Xi$ matter is in fact more deeply bound
than $\Sigma$ or $\Xi$ matter alone. In the following, we will denote the
parameterization responsible for the curves of Fig.\ \ref{fig:earho_hyp}
as model N.

\begin{figure}[tbph]
\begin{center}
\leavevmode
\psfig{file=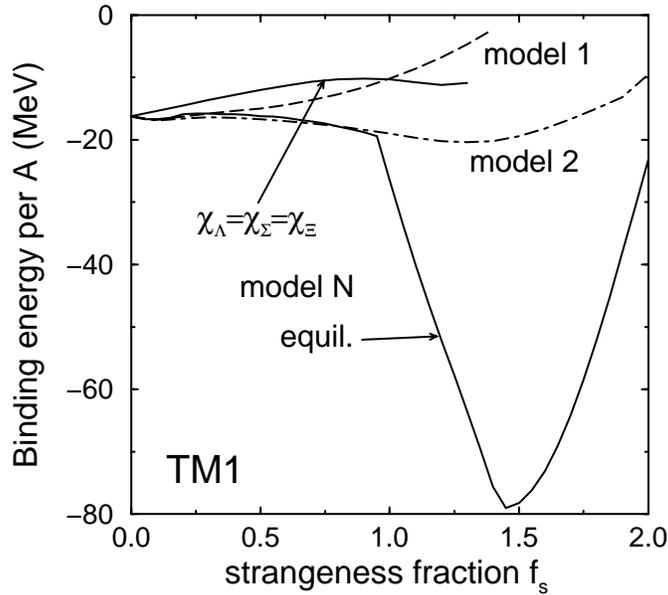,height=0.35\textheight}
\end{center}
\caption{Comparison of the binding energy of SHM per baryon in models
  1 (dash), 2 (dash-dot) and N (solid).
  The upper solid line shows the result for the constrained case of equal
  hyperon fractions ($\chi_\Lambda=\chi_\Sigma=\chi_\Xi$), the lower one
  shows the curve for the correct, unconstrained equilibrium calculation.}
\label{fig:eafs_mn_comp}
\end{figure}

Fig.\ \ref{fig:eafs_mn_comp} shows the binding energy of SHM per baryon in
model N as function of strangeness fraction. For comparison, the curves for
model 1 and 2 from Fig.\ \ref{fig:eafsm3} are also plotted.
We performed two different calculations for model N:
one where the hyperon fractions $\chi_{i}=\rho_{i}/\rho_B$ are held equal by
hand ($\chi_\Lambda=\chi_\Sigma=\chi_\Xi$) as done in Ref.\ \cite{Stoks99b},
and the self-consistent one where the hyperon fractions are determined so as
to ensure chemical equilibrium (denoted as ``equil.'' in the figure).
It is evident from Fig.\ \ref{fig:eafs_mn_comp} that the self-consistent
treatment gives a substantially lower energy, since it is the unconstrained
minimum-energy solution. The disagreement between the results of the two
calculations increases with $f_S$. At $f_S=1$ the difference between the
two curves amounts to nearly $10$ MeV. The curve for model N follows closely
the one for model 2 up to a strangeness fraction of $f_S=0.95$. At larger
values of $f_S$ a deep minimum develops due to the highly attractive
interaction between the $\Sigma$ and $\Xi$ hyperons. Note that an equal
mixture of $\Sigma$'s and $\Xi$'s gives $f_S=1.5$ and the minimum of the curve
is close to that point, i.e.\ $E/A=-79$ MeV at $f_S=1.45$. For larger values
of $f_S$, the curve denoted by N rises again, ending up at $f_S=2$ with the
same binding energy per hyperon as that shown in Fig.\ \ref{fig:earho_hyp}
for pure $\Xi$ matter. The deep minimum around $f_S=1.5$ results from the
deep binding of $\Sigma$ plus $\Xi$ matter which is stronger than seen in
Fig.\ \ref{fig:earho_hyp} for matter composed of either one of these species
separately. This deep minimum structure can only be reached when the hyperon
fractions are allowed to adjust self consistently. Fixing the hyperon fractions,
as done in Ref.\ \cite{Stoks99b}, will always give a curve which is higher in
energy, risking the loss of some important features of the model. In fact, the
curve for the constrained calculation \cite{Stoks99b} ends up at $f_S=4/3$ due
to the particular constraint applied.

\begin{figure}[tbph]
\begin{center}
\leavevmode
\psfig{file=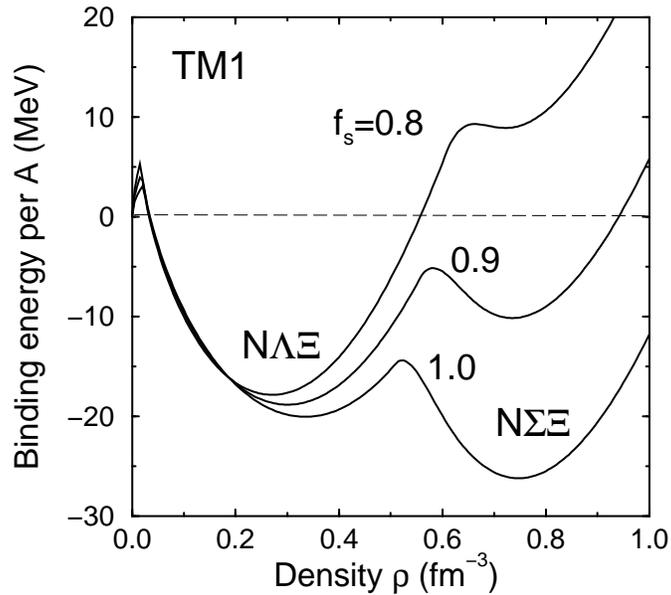,height=0.35\textheight}
\end{center}
\caption{Transition from $N\Lambda\Xi$ matter to $N\Sigma\Xi$ matter in model
  N. A second minimum appears at higher density for a strangeness fraction of
  $f_S=0.8$, becoming more stable for higher strangeness fraction ($f_S=1$).}
\label{fig:earhomn}
\end{figure}

The deep minimum seen in Fig.\ \ref{fig:eafs_mn_comp} emerges due to a second
minimum in the corresponding equation of state at high strangeness fraction,
connected with a
first order phase transition from matter consisting of $N\Lambda\Xi$ baryons
to $N\Sigma\Xi$ baryonic matter. This transition is visualized in Fig.\
\ref{fig:earhomn} where the binding energy is drawn versus the baryon density
for several representative fixed values of $f_S$. For $f_S=0.8$,
there is a global minimum at a baryon density of $\rho_B=0.27$ fm$^{-3}$.
A shallow local minimum is seen at larger baryon density at
$\rho_B=0.72$ fm$^{-3}$. Increasing the strangeness fraction to $f_S=0.9$
lowers substantially the local minimum by about 20 MeV, whereas the global
minimum barely changes. At $f_S=1.0$ this trend is amplified and the
relationship between the two minima is reversed, as the minimum at higher
baryon density becomes energetically lower than the one at lower
baryon density. The system will then undergo a transition from the low density
state to the high density state. Due to the barrier between the two minima,
it is a first-order phase transition from one minimum to the other.

\begin{figure}[tbph]
\begin{center}
\leavevmode
\psfig{file=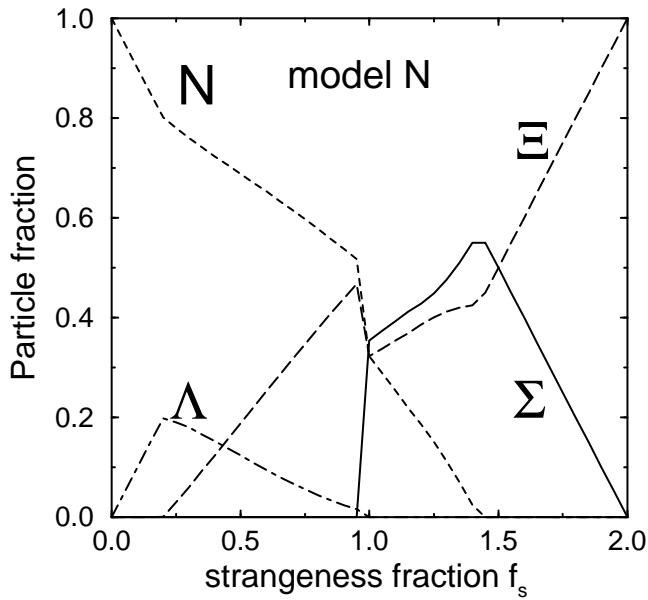,height=0.35\textheight}
\end{center}
\caption{Composition of SHM in model N versus the
  strangeness fraction. $\Sigma$ hyperons appear around $f_S=1$.}
\label{fig:comp_nsc97f}
\end{figure}

Fig.\ \ref{fig:comp_nsc97f} demonstrates explicitly that the phase transition
involves transformation from $N\Lambda\Xi$ dominated matter to $N\Sigma\Xi$
dominated matter, by showing the calculated composition of SHM for model N as
function of the strangeness fraction $f_S$.
The particle fractions $\chi_i$ for each baryon species change as function of
$f_S$. At $f_S=0$, one has pure nuclear matter, whereas at $f_S=2$ one has
pure $\Xi$ matter. In between, matter is composed of baryons as dictated by
chemical equilibrium. A change in the particle fraction may occur quite
drastically when new particles appear, or existing ones disappear in the
medium. A sudden change in the composition is seen in
Fig.\ \ref{fig:comp_nsc97f} for $f_S=0.2$ when $\Xi$'s emerge in the medium,
or at $f_S=1.45$ when nucleons disappear. The situation at $f_S=0.95$ is
a special one, as $\Sigma$'s appear in the medium,
marking the first-order phase transition observed in the previous figure. The
baryon composition alters completely at that point, from $N\Xi$ baryons
plus a rapidly vanishing fraction of $\Lambda$'s into $\Sigma\Xi$ hyperons plus
a decreasing fraction of nucleons. At the minimum of the binding energy curve
in Fig.\ \ref{fig:eafs_mn_comp}, the matter is composed mainly of $\Sigma$'s
and $\Xi$'s with a very small admixture of nucleons.

\begin{figure}[tbph]
\begin{center}
\leavevmode
\psfig{file=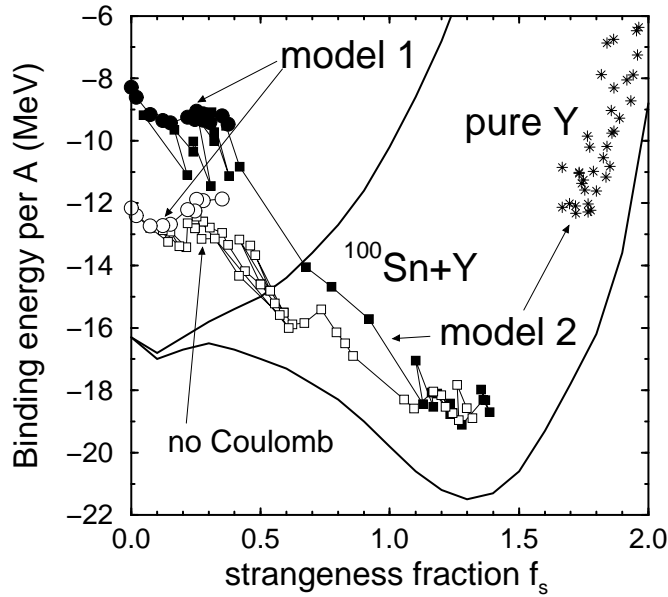,height=0.35\textheight}
\end{center}
\caption{Binding energy of multi-strange finite systems built on a $^{100}$Sn
  nuclear core in models 1 (circles) and 2 (squares, stars), with and without
  Coulomb effects. The curves for bulk SHM (solid lines) are also shown.}
\label{fig:eafsm3_finite}
\end{figure}

Last but not least, we discuss {\it finite} systems of SHM for which the
Coulomb interaction plays a significant role. We also performed calculations
switching off the Coulomb interaction in order to study separately the effects due
to the possibly strong $YY$ interactions. Yet, we will not pursue for finite
systems the implications of the deep minimum found in model N, but rather
stick in the following to the more conservative models 1 and 2. Our choice
for the $\Xi$ potential is $U_\Xi=-18$ MeV, in accordance with the recent
observations \cite{Fukuda98,Khaustov99}. Following the procedure outlined
in \cite{Scha93,Scha94} we start from a ``normal'' nucleus (here $^{100}$Sn,
for example) and add $\Lambda$ hyperons to the system. As soon as the strong
decay reactions $p\Xi^-\rightarrow\Lambda\Lambda$ or
$n\Xi^0\rightarrow\Lambda\Lambda$ are Pauli-blocked, we start adding $\Xi$
hyperons. For each given system of nucleons and hyperons, we check that the two
reactions are blocked also in reverse, so that the whole multi-strange nucleus
is metastable, decaying only via weak interactions. These calculations also
include the effects of the $\rho$ meson field, in order to properly
account for symmetry-energy contributions to the binding energy.

Our results are summarized in Fig.\ \ref{fig:eafsm3_finite}. The solid lines
are the binding energy curves for SHM in bulk for model 1 (upper curve) and
model 2 (lower curve), taken from Fig.\ \ref{fig:eafsm3} for the value
$U_\Xi=-18$ MeV. The filled symbols denote systems where the Coulomb
interaction is included, the open symbols stand for the case where the Coulomb
interaction has been switched off. For model 1, shown in circles, SHM is
slightly more bound than the core nucleus $^{100}$Sn. For the highest
strangeness fraction we found, $f_S=0.375$, the system is bound by $E/A=-9.5$
MeV, compared to $E/A=-8.3$ MeV for $^{100}$Sn. When the Coulomb interaction
is turned off, the curve shifts down by several MeV per baryon.
The chargeless core nucleus $^{100}$Sn is then bound by $E/A=-12.2$ MeV,
whereas the strangeness richest object has now a slightly lower binding energy
per baryon of $E/A=-11.9$ MeV. Hence, the main effect for the increased
binding energy of SHM in model 1 actually comes from Coulomb effects.
The negatively charged $\Xi^-$ hyperons neutralize the positively charged
protons, making SHM more bound than ordinary nuclei.
Still, the binding energies found are well above the curve for bulk matter
due to finite size effects, such as surface tension.

The situation is different in model 2, for which results are shown by squares
in Fig.\ \ref{fig:eafsm3_finite}. Due to the attractive $YY$ interactions in
this model, the binding energy per baryon increases substantially to a value of
$E/A=-19$ MeV at $f_S=1.3$, which is even deeper than the binding energy of
nuclear matter in bulk. This high value of binding is obtained irrespective of
whether or not the Coulomb interaction is included, since for such high values
of strangeness fraction the total charge fraction of the system is close to
zero. Obviously, in model 2 the tremendously increased binding energy of SHM
originates mostly from the attractive $YY$ interactions, and only to a minor
extent from the reduced Coulomb repulsion. Note that the binding energy for
the deepest lying systems in model 2 is close to the values for SHM in bulk
matter which are shown by the lower solid line.
These systems are quite heavy, with mass numbers of about $A=400$ and higher,
so that finite size effects become quite small.

In addition, we also plotted the binding energies of purely hyperonic systems
which consist of $\Lambda\Xi^0\Xi^-$ hyperons solely and, therefore, do not
need to be Pauli blocked in order to keep them metastable \cite{Scha93,Scha94}.
Since these systems can decay only via weak interaction, arbitrary numbers for
the three different hyperon species are allowed. We find that purely hyperonic
objects are bound up to $E/A=-12$ MeV per hyperon. The binding energies,
denoted by stars in the figure, follow the trend of the bulk calculation
curve (solid line).

\section{Summary and conclusions}

In the present work we have calculated the minimum-energy equilibrium
composition of bulk SHM made out of the SU(3) octet baryons $N,
\Lambda, \Sigma$ and $\Xi$, over the entire range of strangeness
fraction $0 \leq f_{S} \leq 2$, for meson fields which generate, within
the RMF model, qualitatively similar baryon potentials to those
generated from the SU(3) extension of the NSC97 potential model
\cite{Stoks99a} using the BHF approximation \cite{Stoks99b}. Our main
results are displayed in Fig.\ \ref{fig:eafs_mn_comp} which shows that
SHM is comfortably
metastable in this model N for any allowed value of $f_{S} > 0$. The
$N \Lambda \Xi$ composition and the binding energy calculated
for equilibrium configurations with $f_{S} \lesssim 1$
resemble those of model 2 in our earlier work \cite{Scha93,Scha94}. The
use of models 2 and N \cite{Scha94,Stoks99a} implies sizable $YY$ attractive
interactions which differ, however, in detail between the two models.
The model of Ref. \cite{Stoks99a} yields particularly attractive
$\Xi \Xi$, $\Sigma \Sigma$ and $\Sigma \Xi$ interactions, but vanishingly weak
$\Lambda \Lambda$ and $N \Xi$ interactions. We remark that this extent of
weakness is in fact ruled out by the little information one has from
$\Lambda \Lambda$ hypernuclei \cite{Dover91} and from $\Xi$-nucleus
interactions \cite{Fukuda98,Khaustov99}. On the other hand, model 2 of
Ref. \cite{Scha94} accounts more realistically for the attractive
$\Lambda \Lambda$ and
$N \Xi$ interactions, but ignores altogether $\Sigma$ hyperons which require
exceptionally strong binding in order to overcome the strong-interaction
$\Sigma B \to \Lambda B$ conversion which in free space releases about 75 MeV.
Yet, all these differences between the two models regarding the relative size
of interactions within $N \Lambda \Xi$ dominated matter hardly matter
when it comes to establishing the stability and binding pattern of this
multi-strange matter. In this sense, SHM is a robust phenomenon. The
metastability of SHM has also been recently confirmed within the modified
Quark-Meson Coupling Model \cite{Wang99}.

The difference between models 2 and N clearly shows up for
$f_{S} \gtrsim 1$, where $\Sigma$'s replace
$\Lambda$'s in model N due to their exceptionally strong attraction to
$\Sigma$ and $\Xi$ hyperons. Figs.\ \ref{fig:eafs_mn_comp},
\ref{fig:earhomn}, \ref{fig:comp_nsc97f} of the present work give
evidence for a phase transition, from $ N \Lambda \Xi $ dominated
matter for $f_{S} \lesssim 1$ to $ N \Sigma \Xi$
dominated matter for $f_{S} \gtrsim 1$, with
binding energies per baryon reaching as much as 80 MeV. This effect has
gone unnoticed in previous works which by constraining the composition
of matter in bulk did not allow for the most general minimum-energy
equilibrium configurations. In contrast, our model 2 produces a much
smoother pattern of binding over the entire range of $f_{S}$, with a
gain of only approximately 5 MeV per baryon (at $f_{S} \approx 1.3$)
for the bulk matter calculation. However, for {\it finite}
multi-strange systems the gain can be considerably bigger, due to
getting rid of most of the Coulomb repulsion for such approximately
charge-neutral systems, amounting to almost 11 MeV per baryon for the
examples of Fig.\ \ref{fig:eafsm3_finite}.

We checked also for the critical strength of the $\Xi$-nuclear potential
below which finite systems of SHM would consist only of nucleons and
$\Lambda$ hyperons in model 1. Of course, the critical value for $U_\Xi$
depends on the size of the system. For a nuclear core of $^{16}$O with 8
$\Lambda$'s filling up the s- and p-shells, $\Xi$'s cannot be added to
the system for a potential shallower than $U^c_\Xi=-13$ MeV. In the case
of $^{56}$Ni, this critical value shifts to $U^c_\Xi=-10$ MeV.
For the $^{100}$Sn nuclear core used to demonstrate finite systems of SHM
in our present calculation (see Fig.\ \ref{fig:eafsm3_finite}), we find
a critical strength $U^c_\Xi=-7$ MeV for which the
$\Xi N \to \Lambda \Lambda$ strong-interaction conversion is barely
Pauli-blocked. However, $U_\Xi$ needs to become more attractive than the
above critical values demonstrate, in order that the corresponding
multi-strange finite systems also remain {\it particle stable}. We remind
the reader that the value $U_\Xi = -18$ MeV used in the figure was designed
to agree with the present phenomenological estimates
\cite{Fukuda98,Khaustov99}. In bulk matter,
$\Xi$-nucleus potentials as repulsive as $U_\Xi=+40$ MeV still admit
bound $\Xi$'s just before SHM gets unbound at $f_S=0.7$.
The reason for this behavior is that the constraint $f_S=0.7$
introduces $\Xi$'s in matter even though their interaction is repulsive.
The Fermi momentum of the $\Lambda$'s become sufficiently high
so that it pays to create some seemingly unfavorable $\Xi$'s in order
to lower the $\Lambda$ Fermi momentum. Note that SHM in model 2 always
contains $\Xi$'s, irrespective of the nature of $U_\Xi$, by virtue
of the attractive underlying $YY$ interactions.

While it is true that the RMF model 2 is a schematic model and is linked
only indirectly to the underlying baryon-baryon interactions, it
is nevertheless constrained by $\Lambda$ and $\Xi$ nuclear phenomenology,
and by the few $\Lambda \Lambda$ hypernuclear species reported to date.
The extrapolation to $YY$ channels which underlie the hyperon potentials
in hyperon matter is more conservative in this model than in model N inspired
by the SU(3) extension of the NSC97 potential model \cite{Stoks99a}.
We emphasize that although the NSC97 model \cite{RSY99} has been tuned up
to reproduce certain characteristics of $\Lambda$ hypernuclei, particularly
its version NSC97f, the predictions elsewhere of these models appear
invariably ruled out by whatever experimental hints one has to date.
In addition to the exceedingly weak $\Lambda \Lambda$ and $N \Xi$
interactions already mentioned above, the NSC97 model overbinds
$\Lambda$ hyperons in nuclear matter ($U_{\Lambda} \sim -38$ MeV)
and gives rise to quite attractive $\Sigma$ nuclear potential
($U_{\Sigma} \sim -20$ MeV) in BHF calculations \cite{Vidana99}, whereas the phenomenology
of $\Sigma^-$  atoms \cite{Mares95} and `hypernuclei' \cite{Dab00}
indicates a much weaker, if not a repulsive, $\Sigma$ nuclear potential.
Furthermore, the NSC97 model gives rise to a sizable $\Sigma$ nuclear
symmetry energy which is opposite in sign \cite{Vidana99} to that of the
earlier NSC89 model \cite{NSC89} and, more importantly,
to that established phenomenologically \cite{Mares95,Dab00,Harada00}.
We therefore suggest that the consequences of the NSC97 model in dense
SHM, as exemplified here, should be taken with a grain of salt. More
dedicated work is required to amend the pitfalls of this model (for a recent
discussion in this direction see Ref.\ \cite{Hal99}).

\section*{Acknowledgments}

JSB thanks RIKEN, Brookhaven National Laboratory and the U.S. Department of Energy
for providing the facilities essential for the completion of this work.
The work of AG is supported in part
by a DFG trilateral grant GR 243/51-2. AG also wishes to thank Larry McLerran,
Sidney Kahana, and John Millener for their hospitality during a supported
visit to the Nuclear Theory Group at the Brookhaven National Laboratory
in February 2000.

\end{document}